\documentclass[useAMS,usenatbib]{mn2e}
\usepackage{amssymb, hyperref, rotating, multirow, fixltx2e, amsmath, graphicx}
\let\ga=\gtrsim \let\la=\lesssim
\def\kms{~km\thinspace s$^{-1}$}     
\def\Msun{~M$_{\sun}$}     
\def\MsunPerYear{\Msun~yr$^{-1}$}

\title[AGN outflows] 
{Active Galactic Nuclei-driven outflows without immediate quenching in simulations of high-redshift disk galaxies}

\author[Gabor et al.]{
J. M. Gabor,$^{1}$\thanks{Email:jared.gabor@cea.fr}
Fr\'ed\'eric Bournaud$^{1}$ 
\\ $^{1}$CEA-Saclay, IRFU, SAp, F-91191 Gif-sur-Yvette, France  
}

\begin{document}


\pagerange{\pageref{firstpage}--\pageref{lastpage}} \pubyear{2013}

\maketitle
\label{firstpage}

\begin{abstract}
We study outflows driven by Active Galactic Nuclei (AGNs) using
high-resolution simulations of idealized $z\sim2$ isolated disk
galaxies.  Episodic accretion events lead to outflows with velocities
$>1000$\kms and mass outflow rates of about the star formation rate
(several tens of \MsunPerYear).  Outflowing winds escape perpendicular
to the disk with wide opening angles, and are typically asymmetric
(i.e. unipolar) because dense gas above or below the AGN in the
resolved disk inhibits outflow.  Owing to rapid variability in the
accretion rates, outflowing gas may be detectable even when the AGN is
effectively ``off.''  The highest velocity outflows are sometimes, but
not always, concentrated within $2-3$~kpc of the galactic center
during the peak accretion.  With our purely thermal AGN feedback model
-- standard in previous literature -- the outflowing material is
mostly hot ($\gtrsim 10^6$~K) and diffuse ($n_H \lesssim
10^{-2}$~cm$^{-3}$), but includes a cold component entrained in the
hot wind.  Despite the powerful bursts and high outflow rates, AGN
feedback has little effect on the dense gas in the galaxy disk.  Thus
AGN-driven outflows in our simulations do not cause rapid quenching of
star-formation, although they may remove significant amounts of gas
over long ($\gtrsim$Gyr) timescales.
\end{abstract}


\section{Introduction} 
Many models of the co-evolution of galaxies and supermassive black
holes (BHs) invoke active galactic nuclei (AGN)-driven winds to
regulate black hole growth and impact the host galaxy
\citep[e.g.][]{ciotti97, haehnelt98, silk98, fabian99, ciotti01,
  wyithe03, begelman05, dimatteo05, sazonov05,
  springel05_mergers_ellipticals, king05, hopkins06_unified,
  robertson06a, somerville08, ciotti10, king10a, king10b,
  dubois10,novak11, teyssier11, debuhr12, dubois12_dual, martizzi12, dubois13b, dubois13}.
Such outflows may help establish correlations between the black hole
and galaxy properties \citep[e.g.][]{kormendy95, magorrian98,
  richstone98, ferrarese00, gebhardt00, mclure02, tremaine02,
  marconi03, treu04, peng06, graham07, treu07, jahnke09, bennert11,
  canalizo12, graham13, kormendy13}.  Powerful AGN winds may help expel sufficient
quantities of gas from their host galaxies to regulate their
star-formation, and in this way suppress the stellar mass growth of
high-mass galaxies.

Observations have begun to probe galaxy-scale outflows thought to be
driven by AGN feedback \citep{tremonti07, feruglio10, alatalo11,
  lehnert11, nesvadba11, rupke11, aalto12, cicone12, cano-diaz12,
  harrison12, maiolino12, cimatti13, forster-schreiber13,
  gonzalez-alfonso13, rupke13, spoon13, veilleux13, cicone14}.  These outflows
are detected in both ionized and molecular gas, and from the local
Universe up to $z\sim2$.  The highest velocities are often found along
lines-of-sight near the galactic nucleus, and these velocities
($\gtrsim10^3$\kms) seem to be higher than those caused by stellar
feedback \citep[e.g.][]{rupke05, martin05, rupke11}.  These high velocities
and central locations are frequently used to argue that the outflows
are powered by AGN rather than supernovae or young stars.  With some
assumptions, mass outflow rates are estimated at $\sim
100$\MsunPerYear or sometimes even higher, or a few times the
star-formation rate \citep[SFR; e.g.][]{cicone14,
  forster-schreiber13}.

Some authors suggest that outflows thought to be AGN-driven are
quenching star-formation in their host galaxies by expelling the gas,
yet most AGN host galaxies are still forming stars.  Most observed
AGNs occur in undisturbed star-forming disk galaxies up to redshift
$z\sim2$, and AGNs are quite common in these galaxies \citep{grogin05,
  pierce07, gabor09, georgakakis09, silverman09, cisternas11,
  kocevski12, mullaney12, rosario12, schawinski12, simmons12,
  juneau13}, although especially luminous or obscured AGNs may be more
likely to be hosted by merging galaxies \citep{bennert08, cales11,
  bessiere12, ramos-almeida12, treister12, juneau13, hopkins13}.  It
remains unclear whether moderate-luminosity AGNs in disk galaxies can
drive outflows or have a major impact on the evolution of their host
galaxies.  Observed outflows attributed to AGN in fact sometimes occur
in galaxies with little or no indications of current AGN
\citep{tremonti07, forster-schreiber13}.

In \citet{gabor13} we presented simulations of gas-rich, isolated disk
galaxies whose supermassive black holes accrete at a high rate.
Accretion is highly variable, with stochastic $\sim10$~Myr episodes of
Eddington-limited accretion.  In the present paper, we show that these
high-accretion episodes generate powerful AGN outbursts, driving hot,
high-velocity outflows from the galaxy.  The outflow rates peak near
the SFR, but the outflows have little overall effect on the gas in the
galactic disk.  We summarize our simulation methods in \S\ref{sec.sims},
describe our results in \S\ref{sec.results}, and discuss caveats and
implications in \S\ref{sec.discussion}.  We conclude with
\S\ref{sec.conclusion}.

\section{Simulations} 
\label{sec.sims} 
The simulations used in this work use the same methods as
\citet{gabor13}.  The simulations are described fully there, and we
summarize them here.

We use a customized version of the Adaptive Mesh Refinement (AMR)
hydrodynamics code \textsc{ramses}\footnote{\url{http://www.itp.uzh.ch/\~teyssier/Site/RAMSES.html}} \citep{teyssier02},
which solves the equations of hydrodynamics and N-body dynamics on a
cubic grid whose cells can vary in size.  We start with an initial
uniform grid of $256^3$ cells in a 50~kpc box and refine to smaller
cells when the mass in a cell exceeds a given threshold, or the gas
Jeans length becomes smaller than $4$ times the cell size.  We allow
refinement up to a maximum resolution (i.e. minimum cell size) of
$\approx 6$~pc.

The code includes models for cooling, star-formation, supernova
feedback, and black hole growth and feedback.  Cooling acts as a sink
term for the thermal energy of the gas \citep[although gas may
  actually gain energy due to heating from a uniform UV
  background][]{katz96,sutherland93}.  We allow gas to cool to 100~K,
except that we use a density-dependent temperature floor at high
densities to ensure the local Jeans length is always resolved by at
least four grid cells \citep{truelove97}.

In gas cells with density $n_H > 100$~cm$^{-3}$ stars form with a
star-formation rate density of $\dot{\rho}_* = 0.01 \rho_{\rm gas} /
t_{\rm ff} \propto \rho_{\rm gas}^{1.5}$ \citep{rasera06}, where
$\rho_{\rm gas}$ is the mass density of gas in the cell, $t_{\rm ff}$
is the free-fall time of the gas, and 0.01 is the star-formation
efficiency parameter \citep{krumholz07}.  New collisionless star particles are
stochastically spawned in eligible gas cells with a probability
appropriate for the star-formation rate.

Ten Myr after each star particle is formed, twenty per cent of its
initial mass explodes as supernovae.  Supernova explosions are modeled
by adding thermal energy to the gas cell in which the star particle
lives.  For each $10$\Msun which explodes as supernovae,
$10^{51}$~ergs are added to the host gas cell's thermal energy.
Cooling is delayed for the heated cell for 2~Myrs to prevent the
injected energy from being radiated away too quickly
\citep[cf.][]{stinson06, teyssier13, perret14}.  This simple stellar
feedback model regulates the structure of the interstellar medium and
suppresses star-formation.  Although the stellar feedback generates
turbulence and thickens the disk, it drives only weak winds that
typically have velocities below the escape speed of the galaxy.
Star-forming clumps throughout the galactic disk launch outflows
locally.  The weakness of the stellar-driven winds in our simulations,
combined with the fact that they are launched throughout the disk,
allows us to isolate the effects of AGN-driven winds in this work.  If
anything, the stellar outflows may have second-order effects on the
long-range propagation of AGN outflows by changing the gas density and
temperature into which AGN-driven shocks propagate.  We control for
such effects in our analysis by comparing simulations with and without
AGN, finding that stellar-driven outflows are typically $\sim
10\times$ weaker than AGN-driven outflows.  More sophisticated stellar
feedback models which include radiation pressure effects \citep[which
  we will examine in future work; e.g. ][]{renaud13} lead to efficient
outflows without AGN, but an ISM structure that is similar to the
simulations we use here \citep{hopkins12_ism, bournaud14, perret14}.

We model a central supermassive black hole as a sink particle
\citep{krumholz04, dubois10,teyssier11}.  We calculate the Bondi
accretion rate $\dot{M}_{\rm BH} = (4 \pi G^2 M_{\rm BH}^2
\rho)/((c_s^2 + u^2)^{3/2})$ \citep{bondi44, bondi52, hoyle39} by
measuring the gas density $\rho$, sound speed $c_s$, and relative
velocity $u$ within a sphere of radius $4 dx$, where $dx$ is the
smallest resolution element.  This accretion rate is capped at the
Eddington limit \citep[cf.][]{eddington16} $\dot{M}_{\rm Edd} = (4
  \pi G M_{\rm BH} m_p)/(\epsilon_r \sigma_T c)$, where $m_p$ is the
proton mass, $\sigma_T$ is the Thomson scattering cross-section, $c$
is the speed of light, and $\epsilon_r=0.1$ is the efficiency with
which accreted mass is converted into luminous energy.  During each
time step the sink particle mass increases and an equivalent mass is
removed from the local gas cells according to the accretion rate.

The amount of mass accreted by the black hole is stored as a variable,
and the corresponding energy may be released as AGN feedback during
coarse timesteps (those that correspond to the coarsest-resolution
cells).  As above, we assume that 10 per cent of the accreted mass is
converted into luminous energy and 15 per cent of the luminous energy
couples to the gas as thermal energy \citep{dubois12_dual}.  If the
accreted mass is sufficient to heat the gas within a radius of $4 dx$
to $10^7$~K, then we inject the energy; if not, then we ``store'' the
accreted mass and check again at the following coarse time step after
additional mass has been accreted.  Storing the accreted mass, though
ad hoc, assures that AGN feedback energy is only injected at
temperatures above $10^7$~K, and thus that the injected energy is not
instantaneously over-cooled and rapidly radiated away \citep{booth09}.
This allows thermal energy injection to develop efficient
pressure-driven outflows.  We also enforce an upper limit to the gas
temperature after energy injection of $5\times10^9$~K to prevent
unphysically high temperatures \citep[see][]{gabor13}.  If the energy
is sufficient to heat the nearby gas above this temperature, we
iteratively expand the injection region until the post-injection
temperature drops below this value.

We construct initial conditions for each isolated disk galaxy with
collisionless star particles in an exponential disk plus bulge (with a
$\sim 20$\% bulge fraction), dark matter particles for the central
regions of a dark matter halo, and an exponential gas disk
\citep{bournaud02}.  We place a black hole with a mass consistent with
the local $M_{\rm BH}-M_{\rm bulge}$ \citep{bennert11} relation at the
galactic center.  The circum-galactic gas is set to a uniform, low
density, $5\times10^{-4}$ of the gas density at the edge of the
galactic disk, or $\sim 10^{-5}$~cm$^{-3}$, and it has a temperature
of $\sim 10^{5.5}$~K.  We allow the initial conditions to relax both
before starting the \textsc{ramses} simulation, and during the first
$100-300$~Myr of evolution with \textsc{ramses} at low resolution to suppress
any spurious density waves \citep[see][]{gabor13}.

In \citet{gabor13} we presented six main simulations: three gas-rich
($\sim 50$ per cent gas fractions) star-forming disks representing
$z\gtrsim 2$ main sequence galaxies, and three low-gas fraction ($10$
per cent) star-forming disks representing low-redshifts galaxies.  We
found little black hole growth and low accretion rates in the
low-redshift simulations.  Our focus in this paper is the
high-redshift, gas-rich regime, where black holes grow substantially
and accrete at the Eddington limit for episodes lasting $\sim
10$~Myr at a time.  

The simulated galaxies have baryonic masses of $1\times10^{10},
4\times10^{10}$, and $16\times10^{10}$\Msun, and we name the runs
M1f50, M4f50, and M16f50 accordingly (where e.g. ``M16'' refers to the
baryonic mass and ``f50'' refers to the gas fraction).  All
simulations begin with gas fractions $>50$ per cent so that, even
after gas consumption due to star-formation, they remain $\approx50$
per cent during the $\sim100$~Myr high-resolution portion of the runs.
In the figures we will show mostly results from simulation M4f50,
which has the most intense black hole accretion history and thus the
most spectacular outflows.  We have re-run this simulation with a
higher frequency of output snapshots to improve the analysis, and the
new version differs from the old only in slight details due to stochasticity in the
simulations.  We will draw conclusions based on all three simulations.
In \citet{gabor13} we also discussed a simulation with no AGN feedback
that is otherwise identical to simulation M4f50.  The two simulations
(with and without AGN feedback) start from the same initial conditions
and undergo the same relaxation procedure.  Here we use the ``no AGN
feedback'' simulation to isolate winds driven by the AGN rather than
supernovae, and to gauge the effect of AGN feedback on the host galaxy.

Using these high gas fraction simulations in \citet{gabor13}, we found
that dense gas clouds (with masses $\sim 10^8 -- 10^9$\Msun) in the
turbulent disk interact and exchange angular momentum, stochastically
converging toward the galaxy center and fueling the black hole
\citep[cf.][]{hopkins_hernquist06, dekel09, ceverino10, bournaud11}.
Black hole accretion is variable on $\sim1$~Myr timescales due to
structure in the ISM, and the dense clouds trigger more prolonged
$\sim10$~Myr episodes of Eddington-limited accretion.  Reaching the
quasar level (up to $\sim 10^{46}$~erg~s$^{-1}$) $\sim 1$\% of the
time, the corresponding AGN luminosities scale with the SFR as in
observations \citep{rosario12}.  The black hole growth -- dominated by
the high-accretion rate episodes -- can explain the black hole growth
required to remain on the $M_{\rm BH}-M_{\rm bulge}$ relation.  Thus
the stochastic migration of dense clouds can fuel an AGN in isolated,
gas-rich disk galaxies.  In the remainder of this work, we study the
outflows generated by such AGN.

\section{Simulation results}
\label{sec.results}

%
%
\subsection{AGN outflows in bursts}
\label{sec.bursts}
\begin{figure*}
\includegraphics[width=168mm]{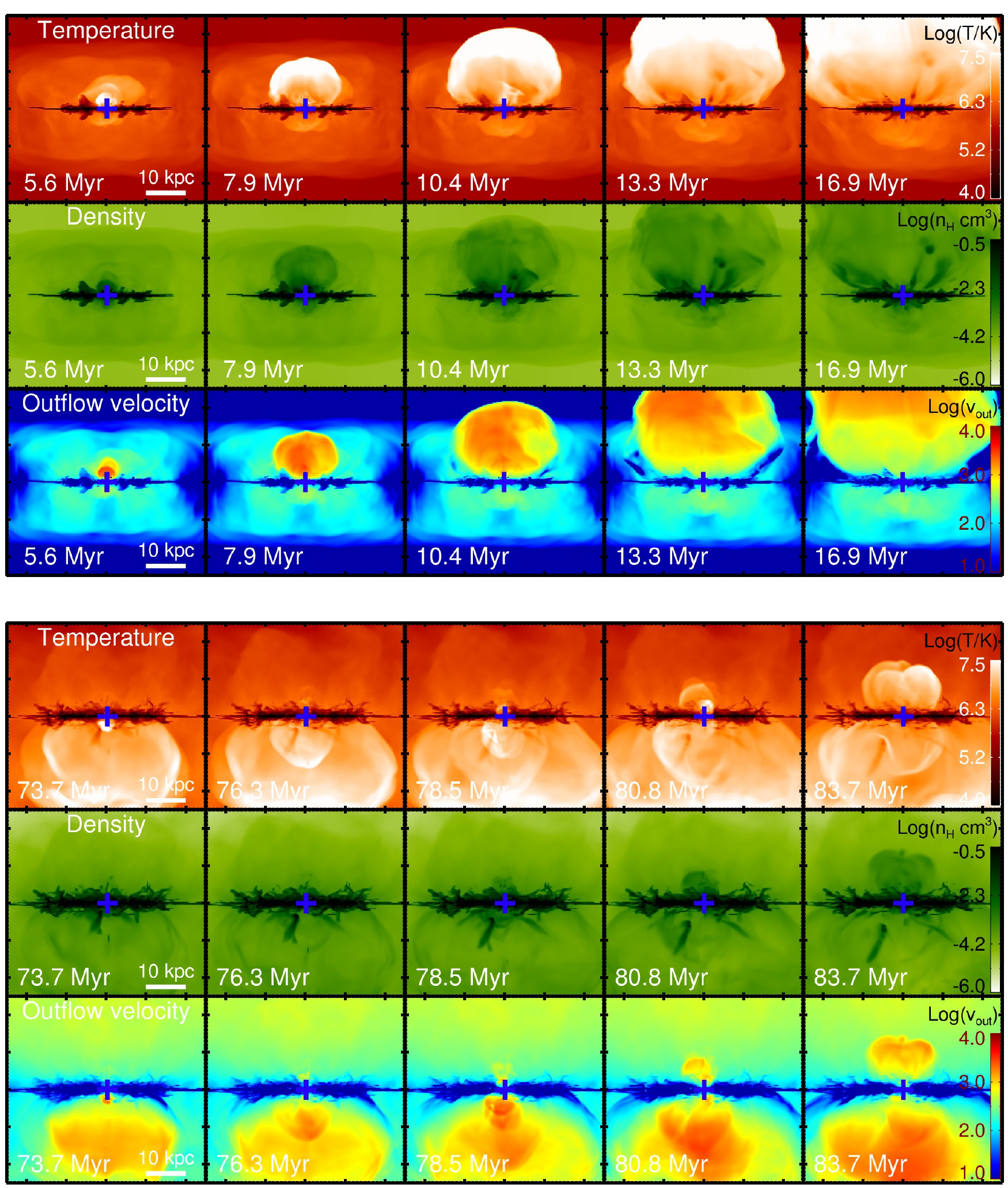}
\caption{Edge-on view of a simulated galaxy disk (M4f50) showing AGN-driven
  outflow bursts.  The top three rows show a single powerful burst in
  a time sequence between 5.6 and 17~Myr, while the bottom three rows
  show multiple bursts between 73.7 and 83.7 Myrs.  For each set, the
  top panels show gas temperature (red scale), middle panels show gas
  number density $n_H$ (green scale), and bottom panels show the
  outflow velocity perpendicular to the disk (rainbow scale).  In both
  examples, an AGN outburst heats gas to $>10^7$~K, generating a shock
  that propagates outward from the disk.  The hot, low-density shock
  front expands at $>1000$\kms, entraining colder denser clouds.  The
  outbursts are frequently asymmetric because dense clouds within the
  disk effectively block the outflows in some directions.  The gas
  disk remains intact.}
\label{fig.snapshots}
\end{figure*}
\begin{figure}
\includegraphics[width=84mm]{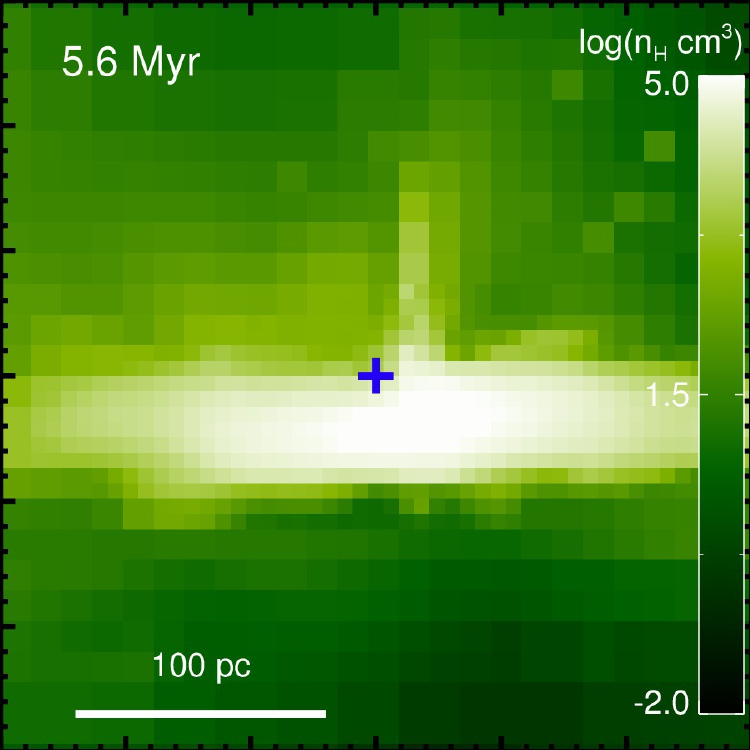}
\caption{Zoom-in on the galactic nucleus during an AGN outburst, from
  an edge-on view as in Figure \ref{fig.snapshots}.  Green-scale
  indicates gas density, and a blue plus sign shows the location of
  the black hole.  The dense gas below the black hole prevents AGN
  feedback from driving an outflow below the plane of the disk.}
\label{fig.blockage}
\end{figure}

%

During high-accretion episodes, the black hole injects powerful bursts
of thermal energy into the immediately surrounding gas, and the
super-heated gas expands and shocks the gas in its path.  Figure
\ref{fig.snapshots} shows two examples of AGN outbursts in simulation
M4f50 through $\sim10$~Myr time-series of snapshots (with time
increasing from left to right).  The figure shows an edge-on view of
the galaxy disk in temperature, gas density, and outflow velocity, all
mass-weighted along the line-of-sight.  The outflow velocity shown is
only the vertical $z-$direction, perpendicular to the disk plane.

The first AGN outburst (top three rows in Figure \ref{fig.snapshots})
expands above the plane of the disk.  The shock front is nearly
semi-spherical.  The outburst does not expand below the disk because
dense structures in the ISM that are close the to black hole block it.
We illustrate the origin of this asymmetry in Figure
\ref{fig.blockage}, where we zoom in on the nuclear region.  The shock
produced immediately around the black hole hits a dense gas cloud
below the disk that absorbs the energy and efficiently radiates it
away.  Above the disk the outflow is free to develop in the
lower-density gas.

The outflow develops primarily as a hot ($>10^7$~K), low-density ($n_H
\lesssim 10^{-2}$) expansion, but the hot gas entrains colder, denser
clouds as well.  One such cloud is visible as a darker clump in the
density map above the disk at 16.9~Myr in Figure \ref{fig.snapshots}.
These swept-up clouds are not as cold and dense as star-forming clumps
in the galactic disk.  We discuss the phase structure of outflows
further in \S\ref{sec.outflow_properties}.

After the outburst launches, the shock wave propagating out of the
disk remains visible for $\sim 10$~Myr, maintaining high velocities
throughout this time.  This timescale is comparable to, and sometimes
longer than, the time during which the black hole accretion rate
remains high.  Since the outflow takes several Myr to propagate to
large distances, it does not fully reflect the rapid variability of
accretion rates.  Thus the AGN-driven outflows may be detectable even
while the AGN is ``off'' (see \S \ref{sec.mass_outflow_rates}).

The second AGN outburst (bottom three rows in Figure
\ref{fig.snapshots}) actually comprises a series of consecutive
bursts.  Each successive burst creates a new shock front expanding in
a shell, leading to an appearance of nested shock shells (e.g. at
78.5~Myr, most obvious in temperature).  Once again the outburst is
asymmetric, with most of the outflow escaping below the disk and
little above it.  Dense clouds are entrained in the hot outflow, and a
bow shock is visible (in temperature at $t=83.7$~Myr) along the edge
of an outflowing cloud below and left of center.  The entrained cloud
moves outward more slowly than the hot medium, and the hot gas forms
the bow shock on the edge of the cloud closest to the AGN.

In both cases, the AGN outbursts apparently leave the galaxy disk
intact -- most of the cold, dense gas appears to remain in the disk,
at least in this edge-on view.  We return to this point in
\S\ref{sec.effects}.

In summary, episodes of high black hole accretion generate brief,
powerful bursts of AGN feedback (see \S\ref{sec.mass_outflow_rates}).
Repeated accretion events lead to repeated, distinct outbursts.  The
AGN-heated, $10^7$~K gas expands away from the gas disk as a shock at
$>1000$\kms, sweeping up cooler, denser clouds along the way.  The
high-density clouds of the ISM are able to halt the shock in its path,
generally leading to asymmetric outflows since the outflow in one
direction may be blocked.  The gas disk of the galaxy is not destroyed
or disrupted by the outbursts.  In the following sections we examine
in further detail the outflow rates, the phase structure of the
outflows, and effects on the galaxy.

%
\subsection{Mass outflow rates}
\label{sec.mass_outflow_rates}
\begin{figure}
\includegraphics[width=84mm]{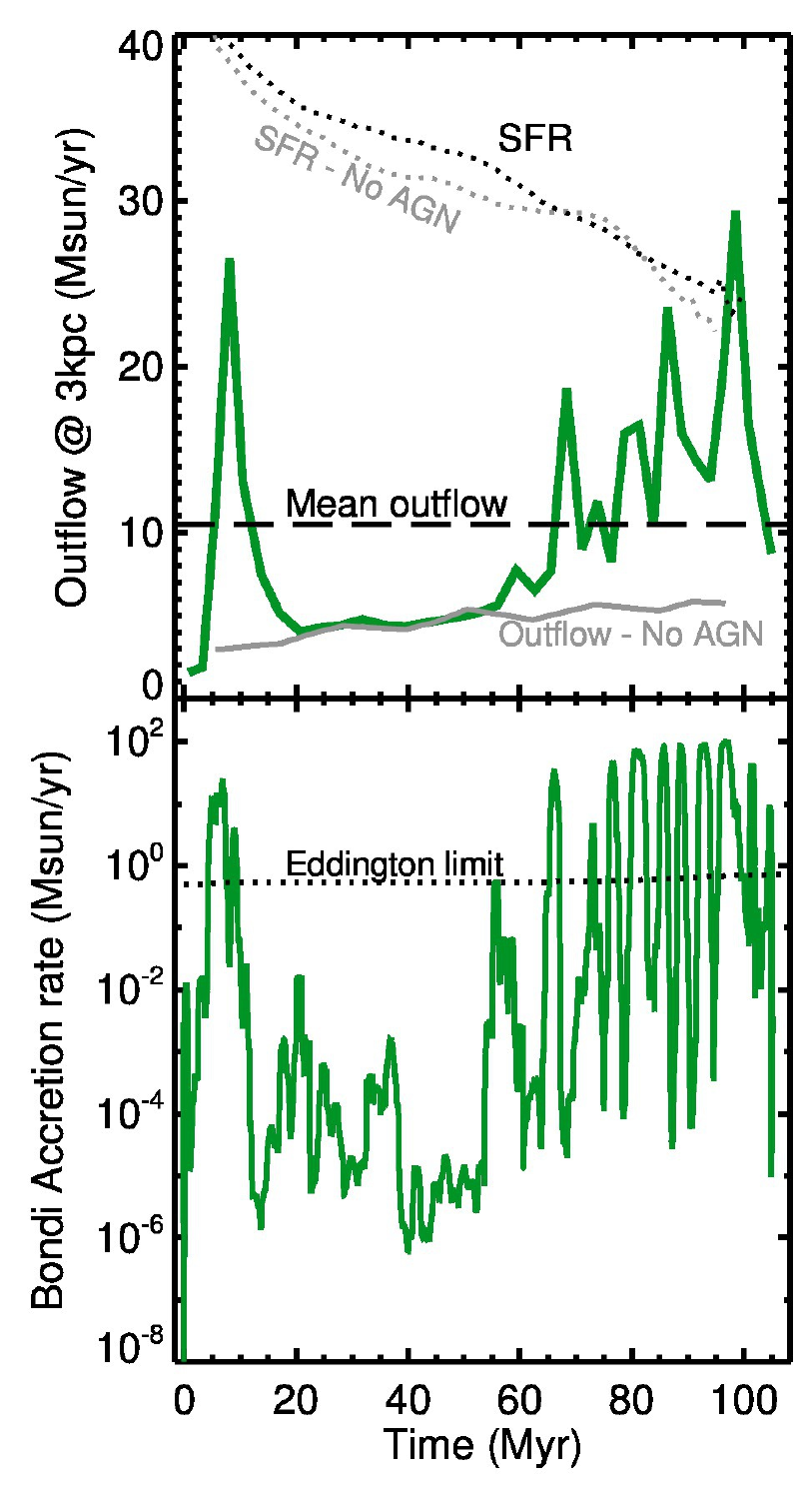}
\caption{{\bf Top:} Mass outflow rate through a plane 3~kpc above (and
  below) the galactic disk, as a function of time (solid green).  A
  dashed horizontal line shows the mean outflow rate during this
  period, and the dotted line shows the SFR for comparison.  {\bf
    Bottom:} Bondi black hole accretion rate as a function of time,
  with the Eddington limit shown as a dotted line.  The actual
  accretion rate of the black hole is capped at the Eddington limit.
  Peaks in the accretion rate lead to outflows in bursts with rates
  comparable to the SFR.  The time-variability of the outflows lags
  that of the accretion rate because it takes $\sim 1$~Myr for
  outflows to propagate from the nucleus to 3~kpc.}
\label{fig.outflow}
\end{figure}
Figure \ref{fig.snapshots} above indicates that AGN feedback generates
hot, high-velocity outflows.  To help understand the broader impact of
AGN on galaxies and their evolution, we measure the mass outflow rates
in our simulations.

We estimate instantaneous outflow rates through planes above and below
the galaxy disk.  Our fiducial planes are 3~kpc above and 3~kpc below
the disk, and they extend to the edge of the box.  For every cell in
the simulation box at a given snapshot, we determine whether its
$z-$velocity would cause it to cross the plane within the next $\Delta
t = 1$~Myr.  If so, and if it is moving outward from the disk, we
count its mass toward the total mass outflow, $\Delta M_{\rm out}$.
We also measure mass inflows but these are typically a factor of $10$
smaller.  The mass outflow rate is then estimated using $\dot{M}_{\rm
  out} = \Delta M_{\rm out}/ \Delta t$.  Different choices of $\Delta
t$ change the outflow rates at the $\sim10$ per cent level. 
  Planes of measurement further from the disk yield slightly lower
  outflow rates -- at $\sim 10$~kpc from the disk, peak outflow rates
  are $\sim 20$\% lower (than at 3~kpc), and they occur later because
  the material takes time to travel to further distances.  We note
  also that some outflowing material moves with a horizontal velocity
  component (i.e. parallel to the galactic disk) that we do not
  include when calculating outflow rates.

We show the mass outflow rates in the top panel of Figure
\ref{fig.outflow}.  As pointed out above, outflows have a bursty
history, following from the episodic accretion history.  For
comparison, we show the accretion rate history in the bottom panel of
Figure \ref{fig.outflow}.  Brief periods of
Eddington-limited accretion generally lead to brief periods of high
outflow rates.  The accretion variability timescale, however, is
sometimes shorter than the timescale for outflows to propagate to our
measurement plane at 3~kpc.  Thus there is not an exact one-to-one
correspondance between intantaneous accretion rate and intantaneous
outflow rate, and outflow rates vary more slowly.  This effect also
allows for substantial outflow rates when the AGN appears to be
``off,'' as in some observations \citep[e.g.][]{forster-schreiber13}.

The outflow rates shown in Figure \ref{fig.outflow} peak around
30\MsunPerYear.  These outflow peaks are similar to the SFR in the
simulation (shown as a dotted line).  The peak outflow rates are also
about two orders of magnitude greater than the black hole accretion
rate.  Due to the bursty accretion history, the outflows reach these
peak rates only for short periods of time.  After the outburst,
outflow rates quickly fall to lower values around $\sim
5$\MsunPerYear.  During active phases of the AGN, the average outflow
rate is about half the peak rate, $\sim 15$\MsunPerYear or the
SFR$/2$.  We also show the mean outflow rate over the 100~Myr
simulation as a dashed line, and it is about 10\MsunPerYear, or
one-third the star-formation rate.  For comparison, our simulation
with no AGN feedback has a relatively constant outflow rate of
$3-5$\MsunPerYear driven by stellar feedback. We also show the
  SFR in the simulation without AGN feedback, which scarcely differs
  from that in the presence of feedback.  Star-formation is not quenched by AGN
  feedback -- we return to this point in \S\ref{sec.effects}.

In our more massive simulated galaxy, M16f50, the black hole accretion
is dominated by a single event near the Eddington-limit.  The outflow
rate peaks around 30\MsunPerYear during this event, with a mean
outflow rate $\sim 10$\MsunPerYear.  This outflow rate is similar to
those found in our medium-mass simulation (cf. Figure
\ref{fig.outflow}), although the SFR is $\sim5$ times higher.  In our
least massive simulated galaxy, M1f50, the outflows are dominated by a
single, longer outburst lasting for $\sim 20$~Myr and peaking at
10\MsunPerYear, or about $2-3\times$ the SFR.  There may be trends
between outflow rates and black hole mass and galaxy disk mass, but
any such trends are too weak to conclusively glean from our small
number of simulations.

In summary, high mass outflow rates follow rapidly after
high-accretion events, as AGN feedback during Eddington-limited growth
episodes drives high-velocity winds.  Since black hole accretion is
rapidly varying, so too are outflow rates.  But the time lag for
outflows to propagate from the black hole vicinity to the measurement
plane (3~kpc) causes the outflow rates to be less variable than
accretion rates.  The peak mass outflow rates are of order the
galaxy's star-formation rate, or in our main example, $\approx
30$\MsunPerYear.  The peak outflow episodes are brief, however, and
outflow rates rapidly return to lower values.  For our simulation with
the most active accretion history, the mean outflow rate is about
10\MsunPerYear, or one-third the star-formation rate.

%
\subsection{Density and temperature of AGN outflows}
\label{sec.outflow_properties}
\begin{figure}
\includegraphics[width=84mm]{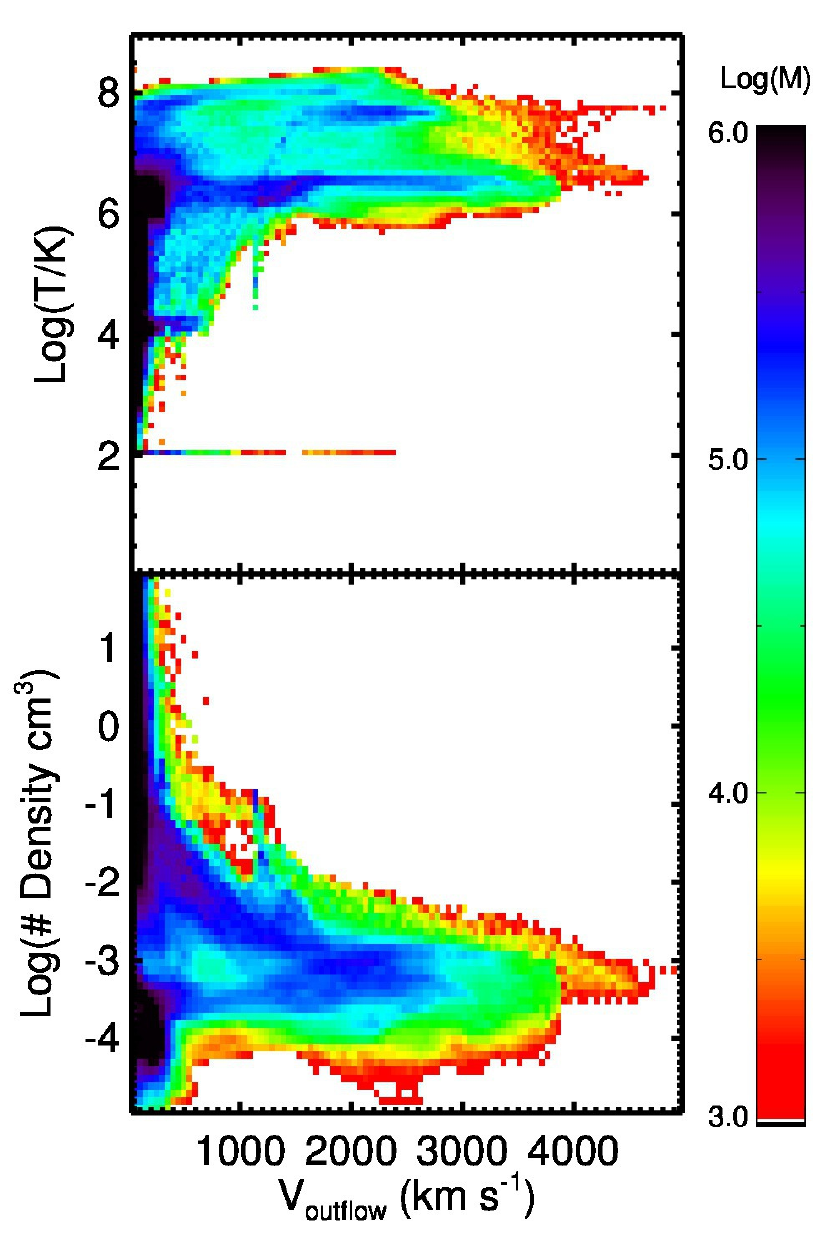}
\caption{{\bf Top:} Distribution of mass in the temperature-velocity
  plane for outflowing gas at a snapshot of simulation M4f50
  ($t=10.4$~Myr).  The color of each pixel indicates the mass at that
  given outflow velocity and temperature, in \Msun.  Outflow
  velocities reach $\approx 5000$\kms, and most of the high-velocity
  gas is between $10^6$ and $10^8$~K.  Some of the outflows up to
  $\sim 2500$\kms are actually at our simulation temperature floor of
  100~K.  These may be cold clouds entrained in the hot
  outflow. {\bf Bottom:} Distribution of mass in the density-velocity
  plane for outflowing gas.  Most of the outflowing gas is
  low-density, below $10^{-2}$~cm$^{-3}$, but some gas at
  $10^{-1}$~cm$^{-3}$ and denser is outflowing at velocities up to
  1000\kms.}
\label{fig.dens_temp_vel}
\end{figure}
Based on Figure \ref{fig.snapshots} we suggested that the outflows are
primarily hot and low-density, but some colder, denser clouds are
entrained in the flow.  The temperature and density structure of
outflows will determine whether and how observations can detect them
-- hot gas may be detectable in X-rays or from ionization lines, while
cold dense gas may be detected via molecular line observations.  The
phase structure also influences how outflows propagate into a
realistic circum-galactic and intergalactic medium.  In this section
we explicitly study the temperature and density of our simulated
AGN-driven outflows.

Figure \ref{fig.dens_temp_vel} shows the gas temperature (top panel)
and density (bottom panel) as a function of velocity for outflowing
gas for one snapshot of a simulation during a powerful outburst.  We
use the snapshot at $t=10.4$~Myr, but the phase structure changes only
slightly as long as the outflows are ongoing.  The color of each pixel
indicates the mass of gas at that pixel's temperature and velocity.
The velocities reach $\approx 5000$\kms.  Most of the high-velocity
outflows have a temperature between $10^6$ and $10^8$~K, but outflows
up to $\sim 1000$\kms have temperatures between $10^4$ and $10^6$~K,
and some outflowing gas is at the temperature floor of our simulation,
$10^2$~K.  Likewise, most of the high-velocity outflowing gas is
diffuse with $n_H < 10^{-2}$~cm$^{-3}$, but some outflowing gas has
densities $n_H > 10^{-1}$~cm$^{-3}$ with velocities $<1000$\kms.
Little or no outflowing gas has densities $>10^2$~cm$^{-3}$, so
molecular outflows would be rare unless the cold clouds in our
simulation could form sub-resolution molecular clouds.

In summary, most of the high-velocity (up to 5000\kms) outflowing gas
is hot and diffuse, with temperatures of $10^6 - 10^8$~K and densities
$10^{-4} - 10^{-2}$~cm$^{-3}$.  Some of the outflowing gas, however,
has temperatures as low as our simulation's temperature floor (100~K), and
densities $>10^{-1}$~cm$^{-3}$.

%
\subsection{Line-of-sight dependence of outflow}
\begin{figure*}
\includegraphics[width=168mm]{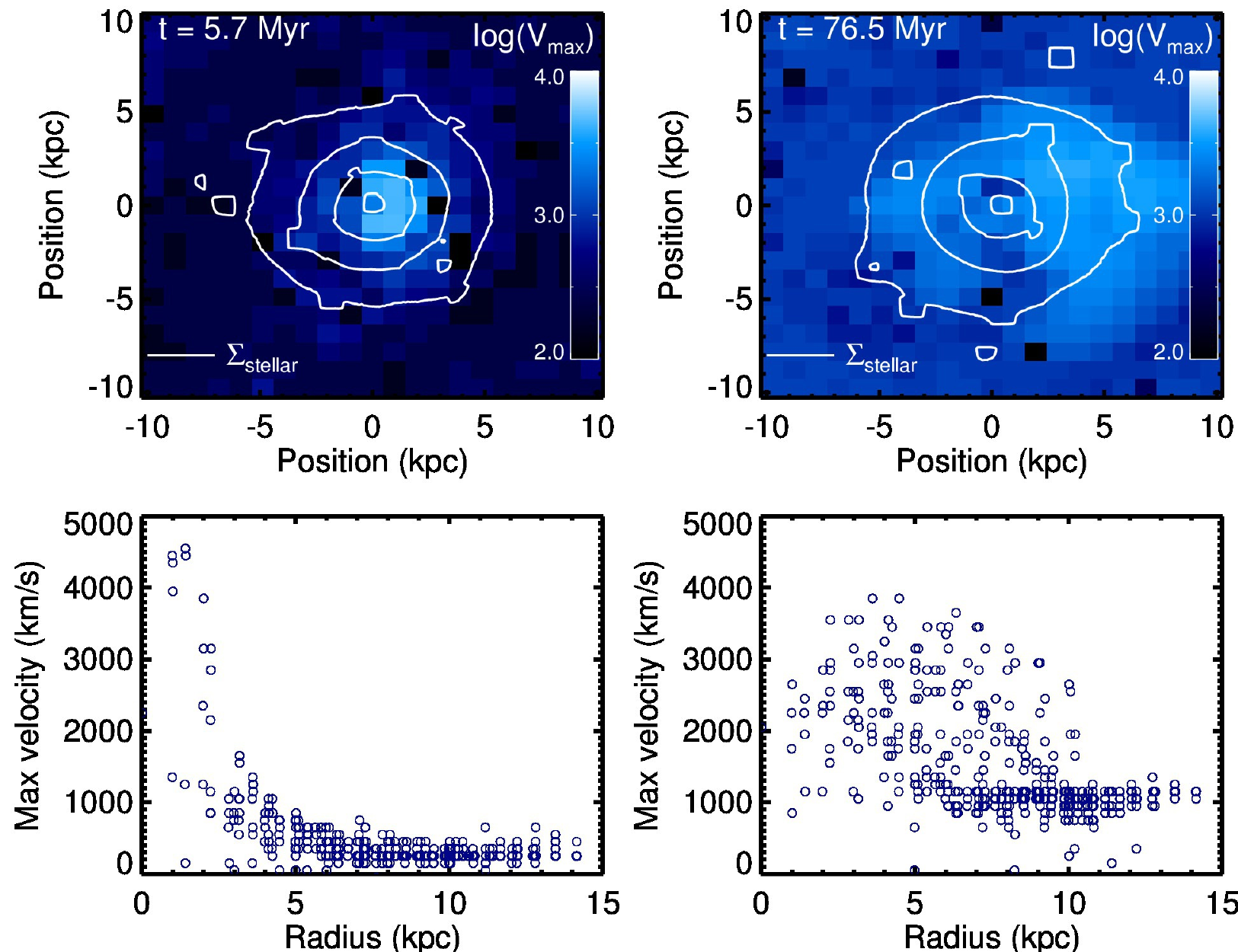}
\caption{{\bf Top Row:} Face-on maps of maximum outflow velocity with
  1~kpc resolution, at $t\approx6$~Myr (left) and $t\approx77$~Myr
  (right).  Contours indicate the distribution of stellar mass of the
  galaxy, from $10^{9.5}$ to $10^{8.0}$\Msun~kpc$^{-2}$ in steps of
  0.5 dex.  {\bf Bottom Row:} Maximum velocity in each pixel from the
  velocity maps above, as a function of the distance of the pixel from
  the galaxy center. During the first AGN outburst ($t\approx6$~Myr,
  left), outflow velocities are highest within $2-3$~kpc of the
  galactic center.  During the repeated outbursts later in the
  simulation ($t\sim70-100$~Myr, right), high outflow velocities are
  seen far from the galactic center.}
\label{fig.maxvel}
\end{figure*}
Observers sometimes attribute galactic outflows to AGNs because
outflow velocities are higher close (in projection) to the galactic
nucleus \citep[e.g.][]{rupke11, forster-schreiber13}.  In this
section, we show that the highest velocities are sometimes, but not
always, concentrated near the galactic center during the peak of an
AGN outburst.


We measure gas velocity profiles along several lines-of-sight through
the galaxy.  The lines-of-sight are set up as a square grid with a
spacing of 1~kpc between each line-of-sight, which is comparable to
the spatial resolution of integral field spectrograph observations of
$z\sim2$ galaxies \citep[e.g.][]{forster-schreiber13}.  Each line-of-sight passes through
the galaxy perpendicular to the galactic disk, emulating a ``face-on''
observation.  We include only gas with temperature $>10^4$~K (that is
also above our temperature floor) to approximate observation of
emission-line gas.  

The top left panel of Figure \ref{fig.maxvel} shows a map of the maximum outflowing gas
velocity at time $t=6$~Myr of simulation M4f50, during the first major
peak of BH accretion.  To limit noise induced by small quantities of
gas, we actually define the maximum velocity along a given
line-of-sight as the highest velocity at which at least 1 per cent of
the gas along that line-of-sight is moving.  The lines-of-sight with
the highest velocities are concentrated near the galactic center,
though the centroid appears somewhat off-center.  

In the bottom left panel of Figure \ref{fig.maxvel}, we show the maximum
line-of-sight gas velocity as a function of projected distance between
each line-of-sight and the galactic center.  The maximum outflow
velocity rapidly drops below $\sim1000$\kms between 2 and 3~kpc from
the galactic center.  Thus during the peak of AGN activity, the
enhanced outflow velocities have an apparent radial extent of
$\sim2-3$~kpc.  

The right panels of Figure \ref{fig.maxvel} present an alternate
  case, during the successive bursts over the last $\sim30$~Myr of the
  simulation.  Here we view the disk from below.  At $t=76.5$~Myr the
  AGN has an Eddington-limited accretion rate, yet high-velocity
  outflows are seen over a wide area of the disk, with velocities
  $>3000$\kms at radii $>5$~kpc from the galactic center.  This occurs
  because the shock shell of a previous AGN burst has expanded
  transverse to the line-of-sight (see bottom panels of Figure
  \ref{fig.snapshots}), propelling fast outflows at large projected
  radii.  In this case it would be difficult to distinguish an
  AGN-driven outflow based on high outflow velocities near the
  nucleus.  We note that an outflowing shell of such large size
  may be expected to interact with a more complex circum-galactic
  medium than that which we model here.

In summary, during the first peak of AGN activity, lines-of-sight
within $\sim 2-3$~kpc of the galactic center show maximum outflow
velocities of several thousand \kms, with lower velocities at larger
radii.  This effect is somewhat sensitive to the timing.  During the
repeated outbursts later in the simulation, outflowing shells expand
to large radii, leading to high-velocity outflows far (in projection)
from the galactic center during subsequent peaks in the accretion rate.

%
\subsection{Effects of AGN feedback on the galaxy}
\label{sec.effects}

\begin{figure*}
\includegraphics[width=6in]{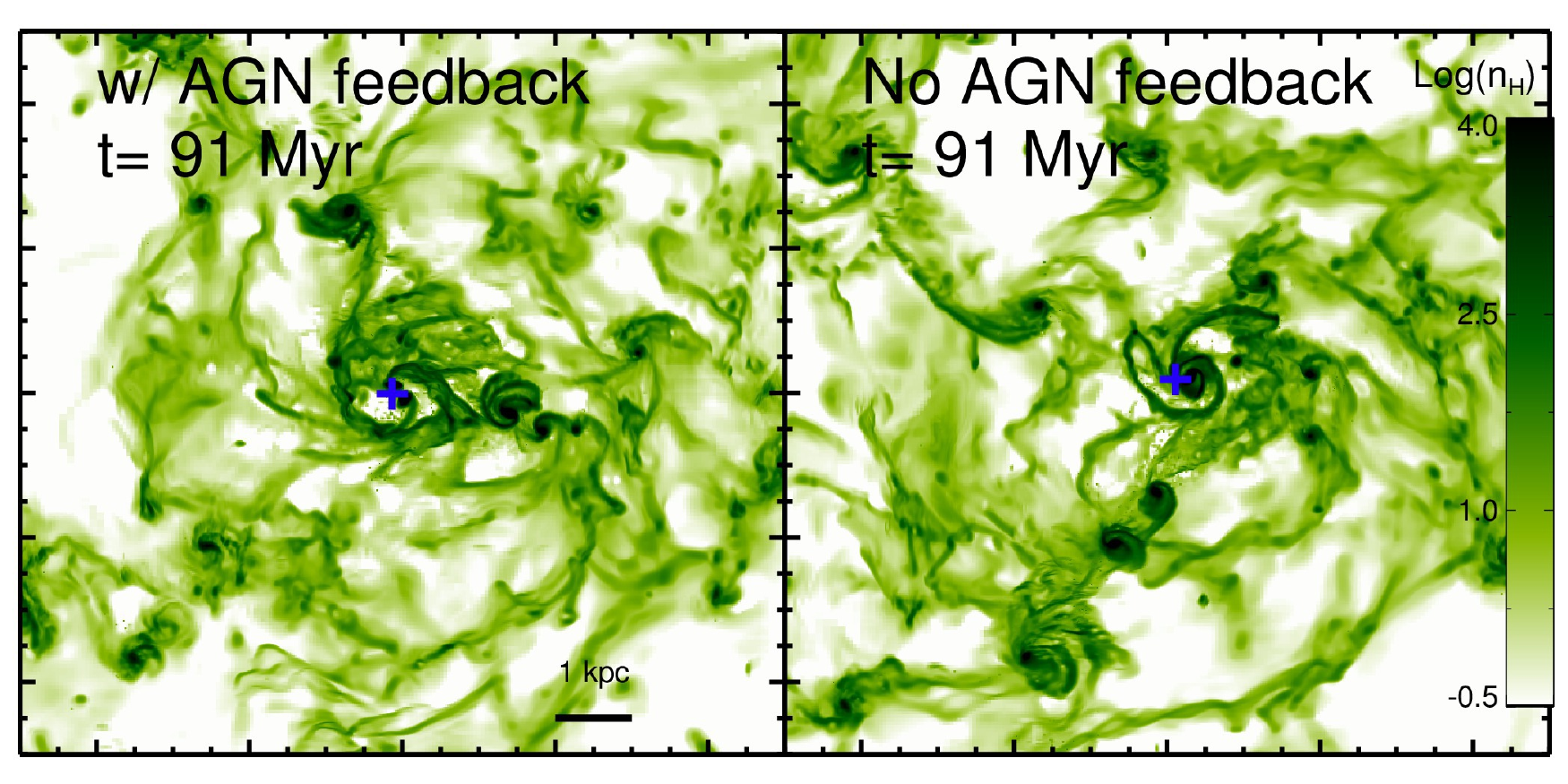}
\caption{Face-on gas density maps of simulations with (left) and
  without (right) AGN feedback, at the same simulation time.  AGN
  feedback changes the details of the gas distribution and kinematics,
  but not the qualitative gas properties.  It affects the evolution of
  individual clouds, but it does not destroy the gas disk.}
\label{fig.faceon}
\end{figure*}
Although AGN outbursts drive peak mass outflow rates near the SFR, and
mean outflow rates of a significant fraction of the SFR (Figure
\ref{fig.outflow}), they appear to leave the galactic disk intact
(cf. Figure \ref{fig.snapshots}).  In this section we show more
explicitly that AGN feedback in our models has little impact on
the star-forming gas disk.

Figure \ref{fig.faceon} shows face-on gas density maps of simulations
with and without AGN feedback at time $t=91$~Myr.  In detail, AGN
feedback alters the evolution of structures within the gas disk.  It
strips gas from the outskirts of individual dense clouds and may
change their dynamics and orbits.  But AGN feedback does not generally
destroy dense clouds, and the galactic gas disk remains intact.  The
gas appears to have similar turbulent structure with filaments and
dense clouds in both cases.

\begin{figure}
\includegraphics[width=84mm]{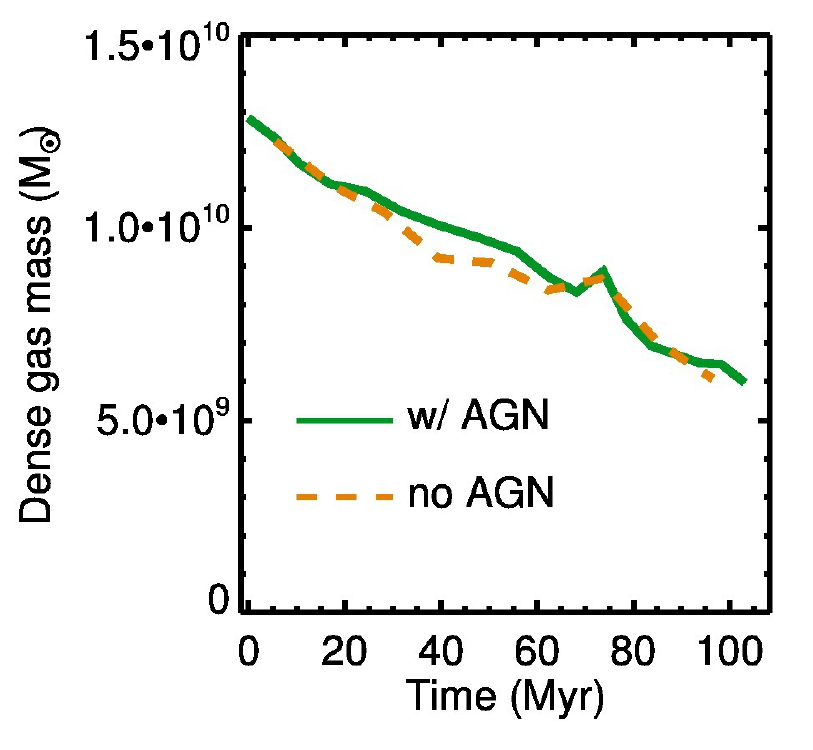}
\caption{Mass of dense gas ($n_H > 100$~cm$^{-3}$) within 5~kpc of the
  galaxy nucleus as a function of time, for simulations with (green
  solid line) and without (dashed orange) AGN feedback.  In both
  simulations the dense gas content declines as it forms stars, and
  AGN feedback makes almost no difference.}
\label{fig.densgas}
\end{figure}

Figure \ref{fig.densgas} compares the evolution of dense gas mass for
the simulation with AGN feedback to the one without AGN feedback.  The
two simulations are identical at time zero.  In each simulation, we
sum the mass of gas above a density of $100$~cm$^{-3}$ within 5~kpc of
the galactic center.  This density is that above which stars may form
in our simulations, and also an estimate for the density required to
form molecular hydrogen \citep{gnedin09}.  The dense gas mass declines
steadily in both simulations due to gas consumption via star formation
and stellar-driven feedback.  Both simulations show a small increase
in dense gas at $t\approx70$~Myr, probably due to a massive clump in
the outer disk moving within 5~kpc.  Remarkably, the simulations with
and without AGN feedback have almost identical dense gas mass
evolution.  The simulation with AGN feedback actually has more dense
gas for some of the time, possibly due to compression from the AGN
outbursts, but the difference is minor.  This similarity in dense
  gas evolution is reflected in the SFRs of the fiducial and ``no
  AGN'' simulations, as shown in Figure \ref{fig.snapshots}.

Why doesn't the AGN feedback strongly affect the dense gas?  When
  the AGN-triggered shock propagates from diffuse gas to denser gas,
  the effective thermal energy increase scales inversely with the gas
  density \citep{mckee75}.  Thus most of the shock energy stays in the
  diffuse phase.  Furthermore, gas cooling rates scale with the
square of density, so that when a shock propagates through a dense
cloud, the added shock energy may be rapidly radiated away
\citep[cf.][]{king10a}.  Within the disk plane, dense gas near the AGN
effectively shields gas in the outer disk from the effects of the
outburst.

In summary, AGN feedback subtly alters details of the gas dynamics in
the central disk regions over time, but it primarily affects diffuse
gas and does not destroy the disk.  The mass of dense gas in the
central 5~kpc is essentially unchanged by AGN feedback.  In upcoming
work we will study how AGN feedback -- including ionizing radiation
from the central source -- affects high-redshift disk galaxy SFRs
(Roos et al. in preparation).
%

\section{Discussion}
\label{sec.discussion}
%
%
\subsection{Model-dependence}
AGN feedback in our simulations drives bursty, high-velocity, hot
outflows with mass rates peaking briefly near the star-formation rate.
Recent work suggests that the details of black hole growth and AGN
feedback can depend quite strongly on the models for these phenomena
\citep[e.g.][]{nayakshin10, ostriker10, debuhr10, novak11,power11, faucher-giguere12,
  angles-alcazar13b, newton13, wurster13, wurster13_compare}.  In this
section we speculate how our results could be model-dependent, without
performing a detailed (and expensive) parameter study.

\subsubsection{Dependence on the AGN model}
The phase structure of outflows shown in Figure
\ref{fig.dens_temp_vel} probably depends on the AGN feedback model.
Our AGN feedback model operates solely through injection of thermal
energy (i.e. by raising the temperature of surrounding gas), and we
only inject the energy at temperatures from $10^7$ to $5\times10^9$~K
\citep[cf.][]{booth09}.  The characteristic speed corresponding to a
thermal temperature of $10^9$~K -- the sound speed -- is $\sim
5000$\kms.  Though this is a reasonable velocity for broad absorption
line outflows commonly observed in quasars \citep{trump06, tombesi10},
we suspect that we could achieve lower-velocity, cooler outflows with
higher mass loading by restricting the injection temperature to lower
values.  That is, one can imagine using the same amount of AGN energy
to drive an outflow with kinetic energy $\sim m v^2$, but increasing
the mass $m$ and decreasing the velocity $v$.  Some feedback models
operate via momentum injection rather than thermal energy injection,
and these might produce colder, denser outflows as well
\citep{debuhr10,choi14}.  

We tested a straightforward alternative AGN model in which the thermal
energy is injected in a volume $5^3=125\times$ larger than our
default.  In this alternative model we still require a minimum
temperature of $10^7$~K, but the peak temperatures around $10^9$~K are
rarely reached. This model produces an outflow phase structure that is
qualitatively similar to the one shown in Figure
\ref{fig.dens_temp_vel}, including similar outflow velocities.  Peak
outflow rates, however, are about $2\times$ the ones in Figure
\ref{fig.outflow}.  This explicitly shows that higher AGN-driven
outflow rates are possible with different AGN models.  Despite
  the higher outflow rates, the effect on the gas disk and its SFR
  remain negligible, at least on short timescales.

 We also tested models with different minimum temperature
  criteria.  A lower minimum temperature allows AGN energy to be
  injected during phases where the accretion rate is relatively low,
  rather than storing the accretion energy for a delayed, more
  energetic burst.  Thus in principle the AGN may release its energy
  in a more moderate trickle rather than in intermittent bursts.
  However, even when the minimum temperature for AGN energy injection
  is $10^5$~K instead of the fiducial $10^7$~K, the burstiness/duty
  cycle, velocities, and phase structure of outflows remain
  qualitatively unchanged.  This is because outflow-driving is dominated
  by Eddington-limited accretion events.  As we showed in
  \citep{gabor13}, the variability of such events is driven primarily
  by structure in the ISM -- dense gas clouds collide with the black
  hole to trigger Eddington-limited events.  During Eddington-limited
  accretion, the minimum temperature criterion is generally moot
  because the energetic output of the AGN is always sufficient to
  surpass the minimum temperature (even $10^7$~K).  Thus the typical
  accretion event that drives an outflow has an injection temperature
  well above $10^7$~K in any case.  So, while outflow rates are somewhat
  sensitive to the volume of the injection region, our results are
  insensitive to the minimum temperature criterion imposed.

In summary, detailed outflow properties such as velocities,
temperatures, densities, and mass outflow rates are somewhat sensitive
to the AGN feedback model.  We expect alternative implementations to
produce colder, slower outflows \citep{choi14}.  On the other hand,
the AGN burstiness, short duty cycle, asymmetric outflow patterns, and
lack of impact on the galactic disk appear to be robust.  Future
simulations can better quantify the robustness of these various
results.

\subsubsection{Dependence on the ISM and stellar feedback models}
The ``burstiness'' of outflows in our simulations arises directly from
the highly variable black hole accretion rates \citep[which is common
  in AGN models; cf. ][]{ciotti10, novak11, hickox13}, which in turn are driven
by structure in the ISM \citep{gabor13, dekel09, ceverino10,
  bournaud11}.  The ISM structure is sensitive to details of cooling
and stellar feedback.  Some simulations model a pressurized ISM which
tends to be smoother and form fewer dense clouds
\citep[e.g.][]{springel03, dimatteo05,
  springel05_mergers_ellipticals}.  Strong stellar feedback, sometimes
involving delayed gas cooling, momentum input, radiation pressure, or
cosmic rays, may also help disrupt dense clouds \citep{genel12,
  hopkins12_clumps}.  In our simulations (with weak stellar feedback)
the AGN outbursts are fed by dense clouds, so models where these dense
clouds are rapidly destroyed could see lower BH accretion rates and
thus weaker outbursts.  Recent simulations similar to those presented
here, but with a sophisticated model for radiation pressure and
ionization due to young stars \citep{renaud13}, generate strong
outflows but appear to leave the clumpy gas structure intact
\citep{perret14, bournaud14}.  Some theoretical work supports the idea
that massive clumps should survive strong stellar feedback
\citep{dekel13, mandelker13}.  Thus the bursty behavior of AGN
feedback should be fairly robust, though the detailed effects of stronger feedback await future work.

\subsubsection{Feedback interactions}
Along with changing the structure of the ISM, strong stellar feedback
can produce outflows \citep{murray05, oppenheimer06, dubois08,
  murray11, hopkins12_winds, uhlig12, salem14}.  Stellar-driven
outflows are (intentionally) weak in our simulations, but the
interplay between stellar-driven outflows and AGN-driven outflows may
be crucial for understanding how gas is expelled from galaxies.
Simulations indicate that different kinds of stellar feedback --
radiation pressure from young stars, and supernovae -- have a
non-linear compounding effect on outflows.  That is, the outflow rates
with both kinds of feedback simultaneously are higher than the sum of
different feedback processes in isolation \citep{hopkins12_winds}.  We
plan to address this interplay in future work.  A related issue is
cosmic inflow and the circum-galactic medium.  Cosmic inflow may
impact the ISM of high-redshift galaxies \citep{genel12_analytic,
  gabor14}, and AGN feedback may interact with and even disrupt cold
inflows \citep{dubois13}.

%
\subsection{Relation to observations}
Our simulations of isolated gas-rich galaxies with AGN-driven outflows
can potentially explain some observations of high-velocity outflows.
The simulations produce outflow velocities $>1000$\kms which are often
used by observers to attribute outflows to AGN.  The peak mass outflow
rates near the SFR of a few tens of \MsunPerYear are
comparable to (though smaller than) those estimated from observations
\citep{cicone14, forster-schreiber13}.  The outflow size scale of
$2-3$~kpc during peak accretion matches observations, though the size
increases during the later stages of the flow in our models.  The
rapid variability of the AGN combined with the time lag of the outflow
may explain observations of high-velocity outflows in the absence of
AGN signatures \citep{diamond-stanic12, forster-schreiber13}.

Other aspects of observed outflows may be difficult to explain with
our model.  The gas in our simulations is probably
hotter than in observations, although such outflowing hot gas may
exist undetected in real galaxies.  Our model does not produce much
cold, dense outflowing gas, while observations indicate that dense
molecular gas ($\gtrsim 100$~cm$^{-3}$) is outflowing at high
velocities along with ionized gas.  This discrepancy could be due to
cooling on scales that our simulation does not resolve.  Although our
peak resolution is $\sim6$~pc, our adaptive refinement strategy only
applies such high resolution to the densest gas in the galactic disk.
Alternatively, the lack of cold dense outflows in our model could be
due to inadequacies in the model used to generate the outflows.

Future work could benefit from synthetic observations of simulated
outflows.  The hot outflows in our simulations would be highly ionized
and emit in X-rays.  Radiative transfer calculations could assess the
observability of various ionization species' emission and absorption
lines and the X-ray continuum.  They could also help calibrate the
connection between direct observables, such as emission line widths
and shapes, and physical outflow properties such as outflow rates.

\subsection{AGN outflows without quenching star-formation}
Some authors interpret observed high-velocity AGN outflows as a
mechanism for quenching star-formation.  In our models, the AGN has
little effect on the gas reservoir in the disk of the galaxy --
  the timescale for AGN outflows to remove all the gas from the disk is
  $\gtrsim2$~Gyr.  If real AGNs are highly variable, as in our
simulations, then time-averaged outflow rates may be significantly
lower than instantaneously observed outflow rates.


Instead of quenching star-formation, we interpret AGN outflows from
disk galaxies as contributing to the overall galactic outflow.
Observations and models suggest that (stellar-driven) galactic outflow
rates of $\gtrsim 2$ times the SFR are reasonable \citep{murray05,
  erb06, oppenheimer06, weiner09, bouche12, dave12, newman12}, but
these do not quench star-formation because galaxies are constantly fed
additional fuel from the IGM.  Indeed, the key physical element
required to produce a galaxy red sequence is not to expel gas from
galaxies, but rather to prevent gas from entering galaxies in the
first place \citep{dekel06, cattaneo06, bower06, croton06,
  hopkins08_ellipticals, gabor11}.  In continuity or equilibrium
galaxy evolution models the intergalactic inflows are roughly balanced
by star-formation and outflows over most of cosmic history:
$\dot{M}_{\rm inflow} \approx \dot{M}_{\rm SFR} + \dot{M}_{\rm
  outflow}$ \citep{bouche10, dave12, genel12_analytic, lilly13,
  dekel13_toy, dekel14_mandelker}.  Based on our simulations, AGN may contribute
a significant (but sub-dominant) fraction of the outflows,
$\dot{M}_{\rm outflow} = \dot{M}_{\rm outflow,stars} + \dot{M}_{\rm
  outflow,AGN}$, rather than quenching star-formation.

Perhaps the story differs for high-mass star-forming galaxies and mergers.
Observations suggest that a high fraction of high-mass $z\sim2$
galaxies show signatures of AGN-driven outflows
\citep{forster-schreiber13}, implying a higher outflow duty cycle.
These galaxies may have more massive bulges and black holes, so at any
given Eddington ratio the AGN is more powerful, potentially driving a
more massive outflow.  Could these observed AGN-driven outflows
ultimately expel all the gas from the galaxy?  Future simulations with
higher bulge and BH masses could help elucidate the situation.

\subsubsection{A toy model of nuclear outflows}
\begin{figure}
\includegraphics[width=3.25in]{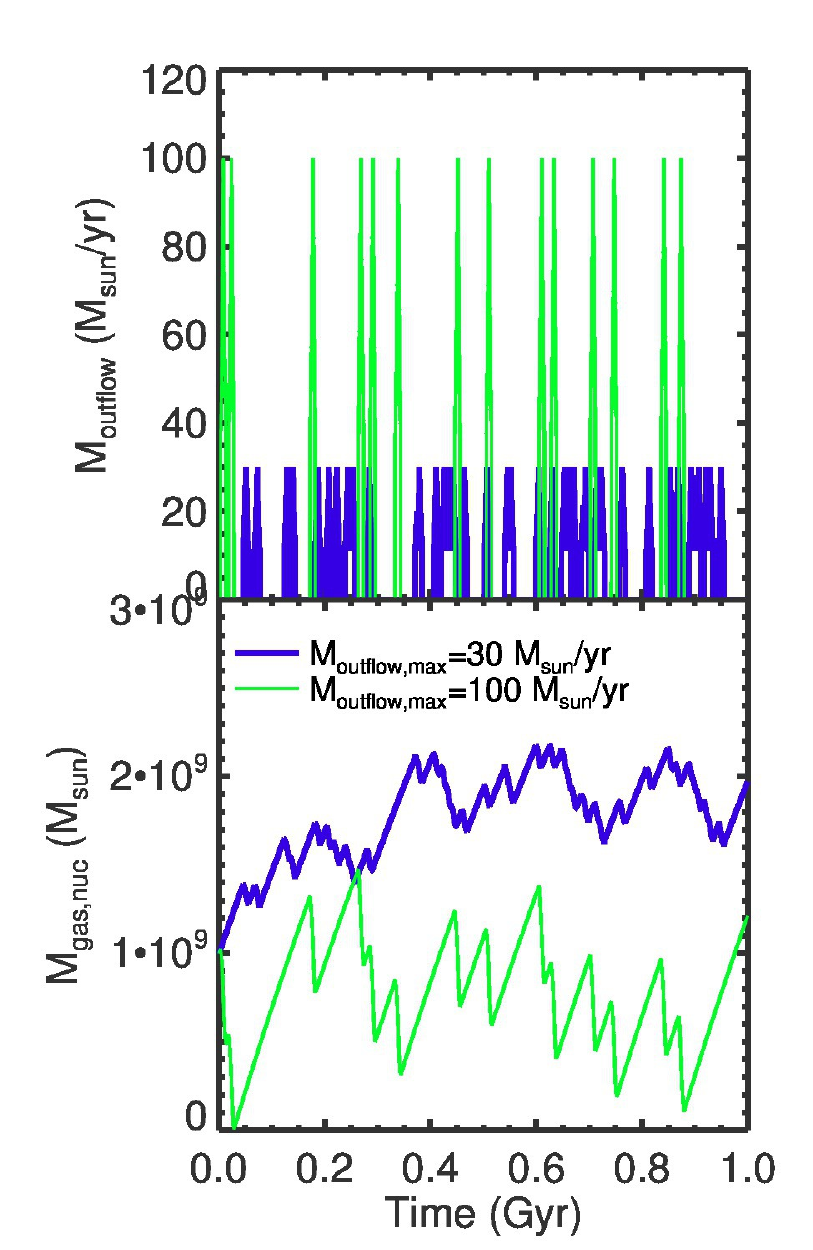}
\caption{Evolution of a toy model for nuclear outflows.  {\bf Top:}
  Mass outflow rates are driven to large values during brief,
  stochastic AGN outbursts.  Here we consider two cases of maximum
  outflow rates: 30\MsunPerYear (blue lines) and 100\MsunPerYear
  (green lines). {\bf Bottom:} The nuclear gas mass fluctates with
  inflows from the galactic disk and outflows driven by the AGN
  outbursts.  Even with high outflow rates, the nucleus spends little
  time devoid of gas.  Note the total galactic gas mass is about
  $10\times$ higher than the nuclear gas mass, and remains roughly
  constant with time -- it is essentially unaffected by AGN feedback.}
\label{fig.toy}
\end{figure}
To illustrate AGN outflows without quenching, we outline here a toy
bathtub-in-a-bathtub model, the key features of which are based on our
simulation results.  The black hole accretes from gas in the nuclear
region, and in our simulations AGN feedback only affects the gas in
the nuclear region, say the central kpc.  Therefore, we treat the
nucleus as a nearly-independent gas reservoir, connected to the rest
of the galaxy disk only through inflows from the disk to the nucleus.
Analytic models \citep{dekel09} and simulations \citep{bournaud11}
indicate that both massive gas clumps and more diffuse clouds flow toward the nucleus of an
unstable gaseous disk galaxy at a rate of $\dot{M}_{\rm
  inflow,nuc} = 0.2 (M_{\rm disk} / t_{\rm d}) (\sigma / V)^2$, where
$M_{\rm disk}$ is the total galaxy disk mass, $t_{\rm d}$ is the disk
dynamical time, $\sigma$ is the velocity dispersion, and $V$ is the
characteristic rotation speed.  For a galaxy like our main simulation,
this gives $\dot{M}_{\rm inflow,nuc} \approx 10$\MsunPerYear.

Gas sink terms from this nuclear region include star-formation, gas
ejected by stellar feedback, and gas ejected by AGN feedback.  The
star-formation rate is $\dot{M}_{\rm SFR, nuc} = M_{\rm nuc} / t_{\rm
  SFR}$, where $M_{\rm nuc}$ is the gas mass in the nuclear region,
and $t_{\rm SFR} = 800$~Myr \citep[appropriate for $z\sim2$ galaxies;
  e.g.][]{genzel10}.  We neglect accretion onto the black hole and
stellar feedback since they are sufficiently small in our simulations
that including them would not substantially change the result.  AGN-driven
outflows are treated as stochastic bursts, with a parametrization
based on the duty cycle in our simulations.  At any given time a new
outburst can begin, with a probability of $0.01 M_{\rm nuc} / M_{\rm
  crit}$, and we choose $M_{\rm crit} = 5\times10^9$\Msun.  Outflow
rates rise for a time $0.5 dt_{\rm burst}$ as 
\begin{equation}
\dot{M}_{\rm outflow} = \dot{M}_{\rm out,max} \exp\left(\frac{t - t_{\rm
      burst} - 0.5 dt_{\rm burst}}{\tau_{\rm burst}}\right),
\end{equation}
 peak at $\dot{M}_{\rm out, max}$, then decline as
\begin{equation}
\dot{M}_{\rm outflow}= \dot{M}_{\rm out,max} \exp\left(-\frac{t - t_{\rm
      burst} - 0.5 dt_{\rm burst}}{\tau_{\rm burst}}\right).
\end{equation}  
Here, $dt_{\rm burst}=10$~Myr is the length of each AGN outburst,
$\tau_{\rm burst}=5$~Myr is an exponential growth/decay timescale for
the outflow rate, $t$ is the time, and $t_{\rm burst}$ is the time at
which the burst began.  Outside of the bursts, AGN-driven outflow rates are
assumed to be zero.  This scheme reproduces the typical frequency
  of bursts and the time-averaged outflow rates of our simulations with
  reasonably good fidelity.

This nuclear reservoir is embedded in the broader reservoir of the
galaxy disk.  This larger gas reservoir gains mass from cosmic
accretion, and loses mass to star-formation, star-formation driven
outflows, and inflows to the nucleus: 
\begin{equation}
\begin{split}
\dot{M}_{\rm gas, disk} = & \dot{M}_{\rm inflow,cosmic} - \dot{M}_{\rm SFR,disk} \\
 &- \dot{M}_{\rm outflow,disk} - \dot{M}_{\rm inflow,nuc},
\end{split}
\end{equation}
where $\dot{M}_{\rm inflow,cosmic}=60$\MsunPerYear is the
intergalactic inflow into the galaxy, $\dot{M}_{\rm SFR,disk}= M_{\rm
  gas,disk}/t_{\rm SFR}$ is the star-formation rate in the disk, and
$\dot{M}_{\rm outflow,disk}=$ is the stellar-driven outflow rate from
the disk.  For simplicity we take $\dot{M}_{\rm
  outflow,disk}=\dot{M}_{\rm
  SFR,disk}$\citep[cf.][]{genel12_analytic}.  With the parameter
choices above, the gas mass (and thus the star-formation rate) of the
disk remains constant over time, as is roughly the case for $z\sim2$
galaxies in equilibrium models of galaxy evolution
\citep[e.g.][]{dave12, genel12_analytic, lilly13, dekel13_toy, dekel14_mandelker}.

Figure \ref{fig.toy} shows the AGN-driven outflow rates and nuclear
gas masses for two example numerical solutions of the above equations
over 1~Gyr -- one case with $\dot{M}_{\rm out,max}=30$\MsunPerYear
(blue), and one with $\dot{M}_{\rm out,max}=100$\MsunPerYear (green).
The top panel shows the mass outflow rate $\dot{M}_{\rm outflow}$ due
to the stochastic, sharp AGN outbursts.  The bottom panel shows the
nuclear gas mass, $M_{\rm nuc}$.  The nuclear gas mass increases due
to the disk inflow, and decreases during AGN outbursts.  Even when the
nuclear region is occasionally cleared of most of its gas (in the case
of high mass outflow rates), the additional gas inflows quickly
rebuild the nuclear reservoir. This model captures the hallmark of
black hole self-regulation -- when AGN outbursts rid the nuclear
region of gas, there is no fuel to trigger an AGN, so more gas is
allowed to build up.

While this toy model ignores some details (e.g. variability in the
inflow rate, ionizing radiation from the central source; Roos et
al. in preparation), it illustrates frequent and episodic strong
outflows {\it without} quenching.  Based on our simulations, the AGN
empties the gas only from the nuclear region, leaving the galaxy disk
intact.  The nuclear region is replenished rapidly due to
instability-driven gas inflows, ultimately triggering additional AGN
and more outflows.  Thus the nucleus persists in a quasi-steady state
where AGN-driven outflows are roughly balanced on long time scales by
gas inflows, despite variability on $\sim$~Myr time scales.  
Over
long time scales, a substantial amount of gas may be ejected from the
galaxy by AGN feedback, but this does not ``quench'' star-formation
because the galaxy accretes new gas.  

This toy model would not apply to some galaxies.  Accretion
  events lasting longer than our typical $\sim10$~Myr may drive
  sustained winds that have greater impact on the host galaxy.
  Furthermore, if most of a galaxy's gas is concentrated near the
  nucleus, as in some merger remnants \citep[e.g.][]{barnes91,
    barnes96,hopkins09_cusps} and blue nuggets \citep{barro13a,
    barro13b, williams14, dekel14_nuggets}, then AGN feedback may be
  more likely to heat or expel that gas and quench most of the
  galaxy's star-formation.  However, even if an AGN (or some other
  process) empties a galaxy of its gas, if that galaxy has a source of
  inflowing gas, it will eventually start forming stars again
  \citep[cf.][]{gabor11}.  This could lead to temporary quenching and
  later rejuvenation.  Generally, nuclei and galaxies with an ongoing
source of cold gas will not be permanently quenched even in the
presence of strong outflows.


\section{Summary and Conclusion} 
\label{sec.conclusion} 
We have studied AGN-driven outflows using a small suite of simulations
of gas-rich, high-redshift isolated disk galaxies.  The main results
from our simulations are:

\begin{itemize}
\item AGN outbursts occur in short, episodic bursts lasting $\sim5-10$~Myr.  
\item AGN feedback drives high-velocity outflows with mass outflow
  rates peaking near the SFR (a few 10's of \MsunPerYear).  The
  time-averaged outflow rate can be $\sim1/3$ of the SFR.
\item Dense gas in the galactic disk directs AGN outflows out of the
  plane of the disk.  The outflows propagate into the circum-galactic
  region as semi-spherical blast waves.
\item AGN outflows are frequently asymmetric, expanding only above or
  below the gas disk, not both. Dense gas clouds vertically offset
  from the central engine can block the outflow in some directions.
\item AGN outflows are predominantly hot and diffuse ($T\sim10^7$~K
  and $n_H\sim10^{-3}$~cm$^{-3}$), but some colder, denser clouds
  ($T\sim10^2-10^4$~K and $n_H \sim 10^{-1}$~cm$^{-3}$) are swept up
  in the hot outflows. These detailed properties of outflows,
    along with outflow rates, are somewhat sensitive to the AGN
    feedback model.  Since the high-accretion events that drive
    outflows have injection temperatures above $10^7$~K in any case,
    our results are \emph{not} particularly sensitive to the to minimum
    injection temperature of our feedback model.
\item During peak BH accretion, the highest outflow velocities are
  sometimes, but not always, visible along lines-of-sight within
  $2-3$~kpc of the galactic center.  In some cases high velocities are
  visible far from the galactic center due to the transverse expansion
  of outflowing shells.
\item AGN feedback has only a weak effect on the detailed dynamics of
  the galaxy's gas.  The mass of dense gas in the disk is almost
  identical for simulations with and without AGN feedback.
\end{itemize}

Beyond these main points, we have speculated on how these results
could be model-dependent, and we have discussed their relation to
observations and the popular idea of AGN-driven quenching.  We
advocate a picture where typical AGN-driven outflows contribute to the
overall galactic outflows, but do not affect the dense gas in the disk
or the SFR.  AGNs fueled by disk instabilities exist in a
quasi-steady state where powerful outflow bursts balance nuclear inflows over
long timescales, without quenching star-formation across the galaxy
disk.


 
 

\section*{Acknowledgements}
We thank the referee, A. Dekel, along with C. DeGraf, for careful reading and many helpful
comments.  We thank R. Teyssier for making \textsc{ramses} available.
We acknowledge support from the EC through grants ERC-StG-257720 and
the CosmoComp ITN.  Simulations were performed at TGCC and as part of
a GENCI project (grants 2011-042192, 2012-042192, and 2013-042192).

\bibliographystyle{mn2e} 

\bibliography{paper}


\label{lastpage}

\end{document}